\documentclass[%
preprint,
 amsmath,amssymb,
 aps,
floatfix
]{revtex4-1}

\usepackage{graphicx}
\usepackage{dcolumn}
\usepackage{bm}
\usepackage{xcolor}
\begin{document}

\title{Coherent generation and control of tunable narrowband THz radiation from laser-induced air-plasma filament}

\author{Xiaoyue Zhou$^1$}
\thanks{The authors contributed equally to this work}
\author{Yuchen Lin$^1$}
\thanks{The authors contributed equally to this work}
\author{Yi Chan$^1$}
\author{Fu Deng$^1$}
\author{Jingdi Zhang$^{1,2}$}
\email{Corresponding author email: jdzhang@ust.hk}
\affiliation{
$^1$Department of Physics, The Hong Kong University of Science and Technology, Kowloon, Hong Kong SAR, China \\
$^2$William Mong Institute of Nano Science and Technology, The Hong Kong University of Science and Technology, Kowloon, Hong Kong SAR, China
}


\begin{abstract}
We report on the proof-of-principle experiment of generating carrier-envelope phase (CEP)-controllable and frequency-tunable narrowband terahertz (THz) radiation from air-plasma filament prescribed by the beat of temporally stretched two-color laser pulse sequence. The pulse sequence was prepared by propagating the fundamental ultrafast laser pulse through a grating stretcher and  Michelson interferometer with variable inter-arm delay. By partially frequency-doubling and focusing the pulse sequence, an air-plasma filament riding a beat note was created to radiate THz wave with primary pulse characteristics (center frequency and CEP) under coherent control. To reproduce experimental results and elucidate complex nonlinear light-matter interaction, numerical simulation has been performed. This work demonstrates the feasibility of generating coherently controlled narrowband THz wave with high tunability in laser-induced air plasma.
\end{abstract}

\maketitle

Optical generation and detection of THz radiation have been of increasing interest owing to its wide range of applications in non-invasive imaging, chemistry, material science and condensed matter physics \cite{tonouchi2007cutting,Basov2011,kampfrath2013resonant}. THz radiation can be generated from transient current in photoconductive antenna, optical rectification in nonlinear crystals and complex dipole radiation from laser-induced air plasma. On the one hand, the two-color laser-induced air-plasma THz source as a probe beam has recently been studied extensively due to the capability to simultaneously deliver THz pulse of ultrashort duration and ultrabroad spectral range spanning far- and mid-infrared frequencies \cite{Cook2000,dai2007terahertz,Kim2007}, bearing unprecedented opportunities for elucidation of fundamental ultrafast dynamics by registering exhaustive spectroscopic information \cite{hu2014optically,lan2019ultrafast}. On the other hand, table-top narrowband THz source as an excitation beam has become highly desirable in order to meet the rapidly growing demands in dynamic investigation and coherent manipulation of novel states in quantum materials, e.g., Floquet engineering \cite{oka2019floquet} and mode-selective excitation \cite{Zhang2014,disa2021engineering}. Many efforts have been put into the coherent generation of narrowband THz radiation but mainly have relied on light-interaction in solid materials, i.e., photo-conductive antenna \cite{Kamada2014}, lithium niobate crystal (LNO) \cite{Uchida2015,jolly2019spectral} and organic compounds \cite{liu2017generation,lu2015tunable}. To date, few attempts have been made to generate tunable narrowband THz wave from gaseous medium, e.g., laser-induced air plasma \cite{Das2010}.

As a common problem faced by narrowband THz and mid-infrared generation via nonlinear wave-mixing process, long-term CEP walk-off may result from environmental factors, e.g., long-term thermal or mechanical drift  \cite{manzoni2010single,yamakawa2020long}. Although there exist many representative demonstrations on coherent polarization control of THz radiation from two-color laser air plasma \cite{manceau2010coherent,dai2009coherent,zhang2018manipulation,PhysRevLett.103.023902}, it has been an under-explored prospect to generate narrowband THz radiation with coherently controllable CEP and center frequency, a desirable feature for phase-stabilized THz radiation to set firm grounds for mode-selective and phase-sensitive creation and manipulation of novel states in quantum materials \cite{matsunaga2014light,yuan2022revealing}.

In this letter, we demonstrate the proof-of-principle experiment for the generation of narrowband THz radiation with both frequency and CEP tunability from an air-plasma filament under the coherent drive of a two-color pulse sequence, which is prepared by the chirped-pulse beating method\textemdash a pulse-shaping technique extensively used in table-top \cite{chen2011generation, uchida2015time, Weling1994} and accelerator-based  \cite{Bielawski2008,Ungelenk2017} tunable narrowband THz source by exploiting nonlinear crystals and relativistic free-electron bunches, respectively. Taking the approach, a tunable low-frequency intensity modulation can be imprinted on the stretched Gaussian intensity profile, of which the duration can continuously vary in accordance with the change in linear chirp of the carrier wave, i.e., second order dispersion \cite{weling1996novel,Kamada2014}. Here, considering an optical pulse with a center frequency of $\omega_0$ and a linear chirp rate $\beta$ acquired by traversing a grating stretcher, its instantaneous frequency follows $\omega(t)=\omega_0+\beta t$. The chirped pulse is subsequently split into halves upon entry of the Michelson interferometer so as to create two replicas at a variable time delay $\tau$ determined by the optical path difference. Instantaneous frequency of individual replicas follow $\omega_1(t)=\omega_0+\beta (t-\tau)$ and $\omega_2(t)=\omega_0+\beta t$, respectively. Therefore, a constant differential frequency, i.e., beat note $\Omega=\beta\tau$, makes its appearance to modulate intensity profile of the superimposed fundamental waves. Such linear dependence on inter-pulse delay $\tau$, in turn, provides us convenient access to controlling the modulation frequency $\Omega$. When combined with the strong-field-induced air-plasma filamentation process, it is the interferometry-based pulse shaping configuration that forges a pulse sequence to impose a background periodic modulation to both the violent ionization of air molecules and instantaneous coherent motion of the resultant free electrons, favorable for coherent emission of quasi-monochromatic THz waves at the frequency $\Omega$.

\begin{figure}[htbp]
\centering
{\includegraphics[width=\linewidth]{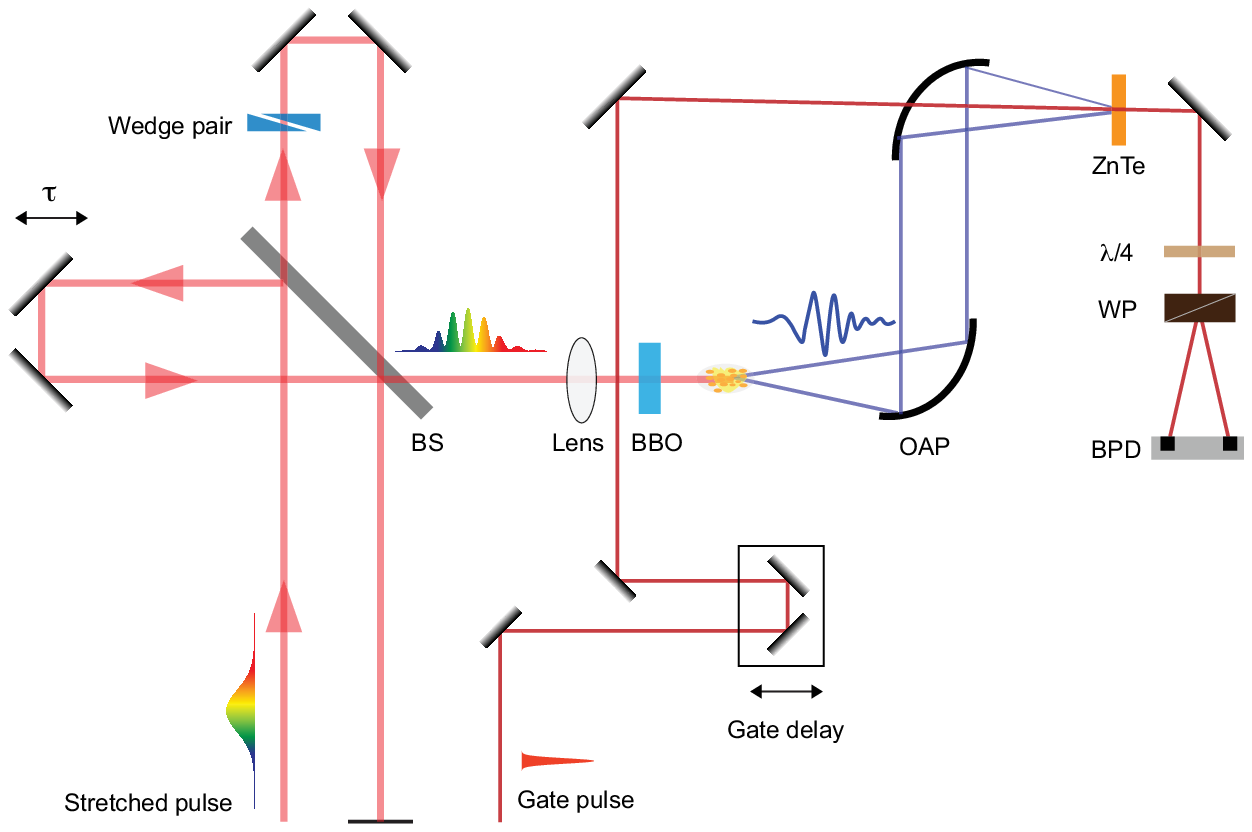}}
\caption{The schematic diagram of experimental setup. BS: non-polarizing beam splitter, OAP: off-axis parabolic mirror, WP: Wollaston prism, BPD: balanced photo-detectors}
\label{fig1}
\end{figure}

\begin{figure}[ht!]
\centering
{\includegraphics[width=\linewidth]{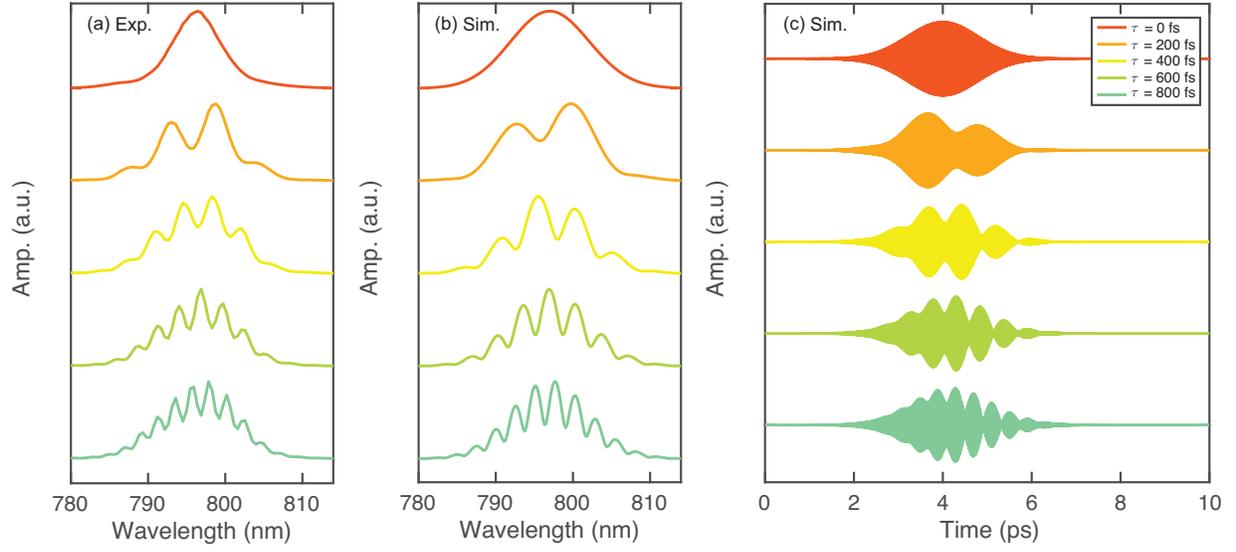}}
\caption{(a-b) Experimental and numerical results showing spectra of fundamental wave at the Michelson interferometer exit with increasing inter-pulse delays $\tau$ (top to bottom). (c) Reconstructed temporal
intensity profiles of the pulse sequence.}
\label{fig2}
\end{figure}

The schematic diagram of our experiment is illustrated in Fig.\ \ref{fig1}. We took the output of a 1 kHz Ti:sapphire amplifier laser system emitting near-IR optical pulses (800 nm in center wavelength, 2 mJ in energy and 100 fs in duration). The chirp level could be readily introduced by varying the grating pair separation of the pulse stretcher. To effectively generate narrowband THz radiation, the pulse duration was stretched to 1.2 ps corresponding to a chirp parameter $C=z\beta_{2}/T_0^2=12$, where $z$ is the deviation in grating position from configuration that produces transform-limited pulse, $\beta_2$ the group velocity dispersion (GVD) parameter and $T_0$ the transform-limited pulse duration. A Michelson interferometer split the stretched optical pulse into two identical pulses at relative delay $\tau$, variable by altering the arm-length difference of the interferometer. Upon recombining and interfering the two at the exit port, a pulse sequence with a quasi-sinusoidal modulation at the beat frequency $\Omega$ could be formed in the time domain. The effectiveness of such modulation is seen at its clearest when being Fourier transformed into the frequency domain, i.e., to verify a periodic modulation to the Gaussian spectrum of individual constituent pulses due to spectral interference effect, Fig.\ \ref{fig2}(a). These spectral features can be easily repeated by taking into consideration the temporal interference at the experimentally set value of inter-pulse delay $\tau$, Fig.\ \ref{fig2}(b). The agreement of the measured and calculated spectra attests an excellent coherence necessary for successful pulse shaping in our experiment and, in turn, allows for a highly reliable reconstruction of pulse sequence in the time domain, Fig.\ \ref{fig2}(c). The resultant pulse sequence served as the fundamental wave (FW) and was focused with its second harmonic (SH) to create the air-plasma filament. Considering excessive duration of the stretched pulse, second harmonic generation (SHG) efficiency for the pulse sequence is moderately lower than that for transform-limited pulse due to reduced field strength. To mitigate the issue, we used a 500 $\mu$m thick BBO crystal for optimal SHG efficiency that presumably scales quadratically with the crystal thickness. A pair of off-axis parabolic mirrors was used to image the emission from air-plasma filament onto the electro-optic sampling (EOS) crystal ZnTe for detecting the time-domain waveform of THz radiation with a transform-limited near-IR gate pulse. To filter out the fundamental and higher-frequency components, a high-resistivity silicon wafer (not shown in the figure) was placed after the plasma source. The entire setup was continuously purged with dry air to eliminate the ambient water vapor absorption.

\begin{figure}[ht!]
\centering
{\includegraphics[width=\linewidth]{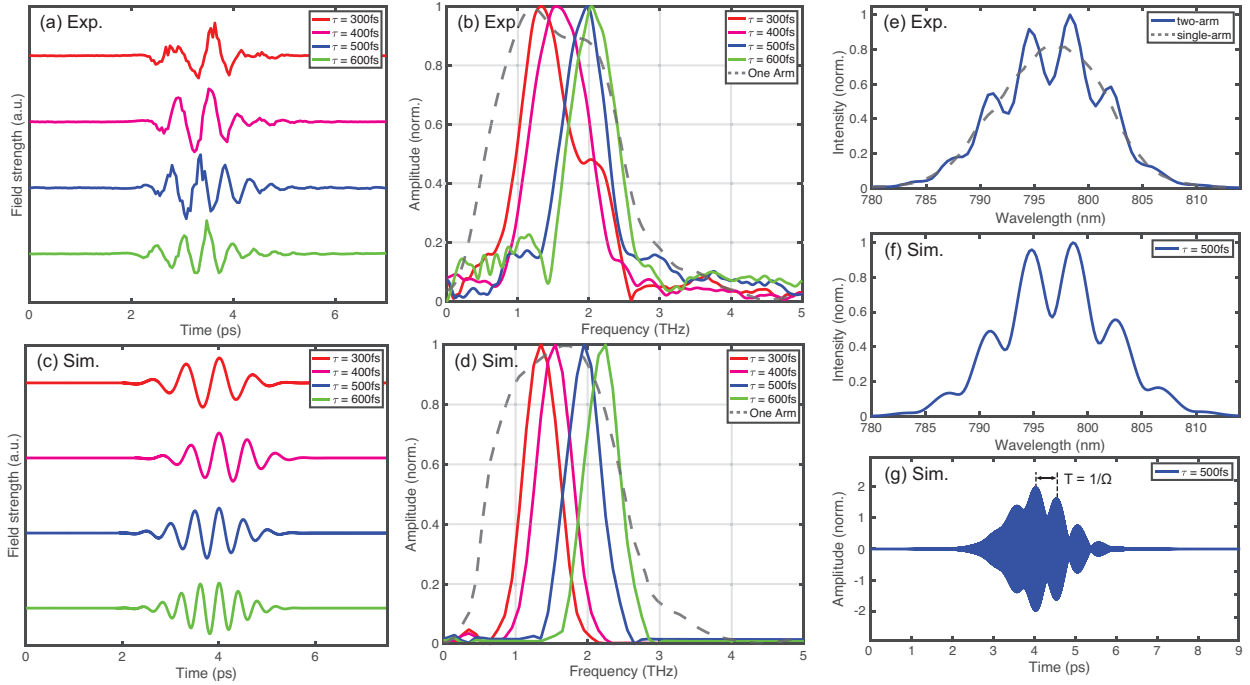}}
\caption{(a-b) Time-domain signals and corresponding intensity spectra of the THz radiation at four representative delays $\tau$ obtained from experiment. The dashed lines are spectra of broadband THz emission from uncontrolled input pulse. (c-d) Numerical simulation on time-domain signals and magnitude spectra of THz radiation. (e) Measured spectrum of the input pulse sequence at delay $\tau = 500$ fs (solid line) and the single pulse (dashed line). (f) Simulated spectrum and (g) reconstructed temporal profile of the pulse sequence.}
\label{fig3}
\end{figure}

We performed a series of measurements on the time-domain THz signal at various inter-pulse delay $\tau$ set by interferometer (Fig.\ \ref{fig3}). In contrast to the typical broadband THz wave radiated from air plasma using uncontrolled two-color laser pulse, the THz signal generated by periodically modulated pulse sequence exhibits multi-cycle oscillatory characteristics in the time domain, signifying a pronounced bandwidth reduction. The linearity between center frequency of the narrowband THz and inter-pulse delay is best demonstrated in the range of 300 to 600 fs in $\tau$, upon which the adjustment can continuously tune the center frequency from 1.28 to 2.02 THz with a minimal bandwidth of 0.725 THz at $\tau=$ 500 fs. In order to optimize the output power of the narrowband THz radiation, the spatial overlap between the twin pulses was aligned to be perfectly collinear, and the BBO crystal was oriented to achieve a FW-to-SHG power ratio optimal for THz generation. However, in this scheme, it is not unexpected to observe the limited tuning range in frequency and deformed narrowband THz waveform under extreme circumstances. In particular, when $\tau$ is large, two adverse effects become the most pronounced, one being the strong phonon absorption in detection crystal and the other being the compromised ionization efficiency and stability in a filament under the drive of a stretched laser pulse with a blunted peak intensity.

As a necessary supplement to the experimental results, we perform numerical simulations to obtain insight into the microscopic dynamics of the emission process. According to the photocurrent model \cite{Kim2009}, the superimposed two-color laser field effectively provides an asymmetric field, triggering the tunneling ionization of electrons in the nitrogen molecues and prescribing a net drift momentum to free electrons. The pertinent quasi-DC motion of free electrons takes place at sub-picosecond timescale and, therefore, radiates at THz frequencies. The tunneling ionization process is predictable by Ammosov-Delone-Krainov (ADK) model \cite{adk}, and the instantaneous current can be described by the expression:
\begin{equation}
    \delta J(t)=\frac{e^2}{m_e}\delta N\int_{-\infty}^t (E^{total}_\omega(t')+E^{total}_{2\omega}(t'))dt',
\end{equation}
where $e$ is the elementary charge of electron, $m_e$ the mass of free electron, $\delta N$ the number of ionized electrons in time interval $[t, t+dt]$. The total current density at a given instant $t$ can be expressed as an integration of $\delta J$ at all prior instants: $J(t)=\int_{-\infty}^t \delta J(t')$. The THz radiation $E_{\text{THz}}$ can be related to the drift current density of electrons $J$ by $E_{\text{THz}}\propto dJ/dt$. In our experiment, the interferometer-prepared laser field follows the form $E^{total}_\omega(t)=E_\omega(t)+E_\omega(t-\tau)$, where $E_\omega(t)$ and $E_\omega(t-\tau)$ denote the electric field of the chirped fundamental wave and its delayed replica, respectively. Note that the envelope of $E^{total}_\omega$ carries a beat note oscillating at a much lower frequency than that of the fundamental wave ($\Omega\ll\omega$). From the equation above, it follows that the concomitant ionization and radiating process repeat at a constant time interval $T$ defined by the inverse of beat note $\Omega$. This periodic modulation to the picosecond ionization event is in equivalence to a "multiple-slit" interference in the time domain. As a result, a quasi-monochromatic electromagnetic wave at frequencies in accordance with the beat note can be emitted and detected in the far-field measurement, shown in Fig.\ \ref{fig3}(a). Taking $\tau$ at 500 fs as an example, we first measured the spectrum of the pulse sequence (Fig.\ \ref{fig3}(e)) and repeated the interference fringes in simulation (Fig.\ \ref{fig3}(f)), from which we could reconstruct the waveform of the pulse sequence in time domain, shown in Fig.\ \ref{fig3}(g). The beat frequency $\Omega$ can subsequently be calculated by taking the inverse of the period of slow amplitude modulation on the pulse sequence, e.g., 1.99 THz with $\tau$ being set to 500 fs. Remarkably, such beat note imposed on the fundamental wave and its direct impact on the output characteristics, i.e. center frequency and transient bandwidth, is witnessed by the time-domain waveform of the resultant THz radiation. Numerical simulation based on photocurrent model is validated, in full, by its excellent agreement with experiments in far-field THz radiation under various delay $\tau$, see Fig.\ \ref{fig3}(c). The efficacy of the approach reported herein is further verified by comparison of experimental and numerical results in the frequency domain, Fig.\ \ref{fig3}(b) and (d), demonstrating a wide tuning range in center frequency $f_C$ in reaction to a varying inter-pulse delay $\tau$. Note that experiments agree well with simulations only when $f_C < 2 \text{ THz}$, whereas discrepancies emerge in case $f_C>2 \text{ THz}$ which is an expected trend when considering the above-mentioned limitations. 

\begin{figure}[ht!]
\centering
{\includegraphics[width=\linewidth]{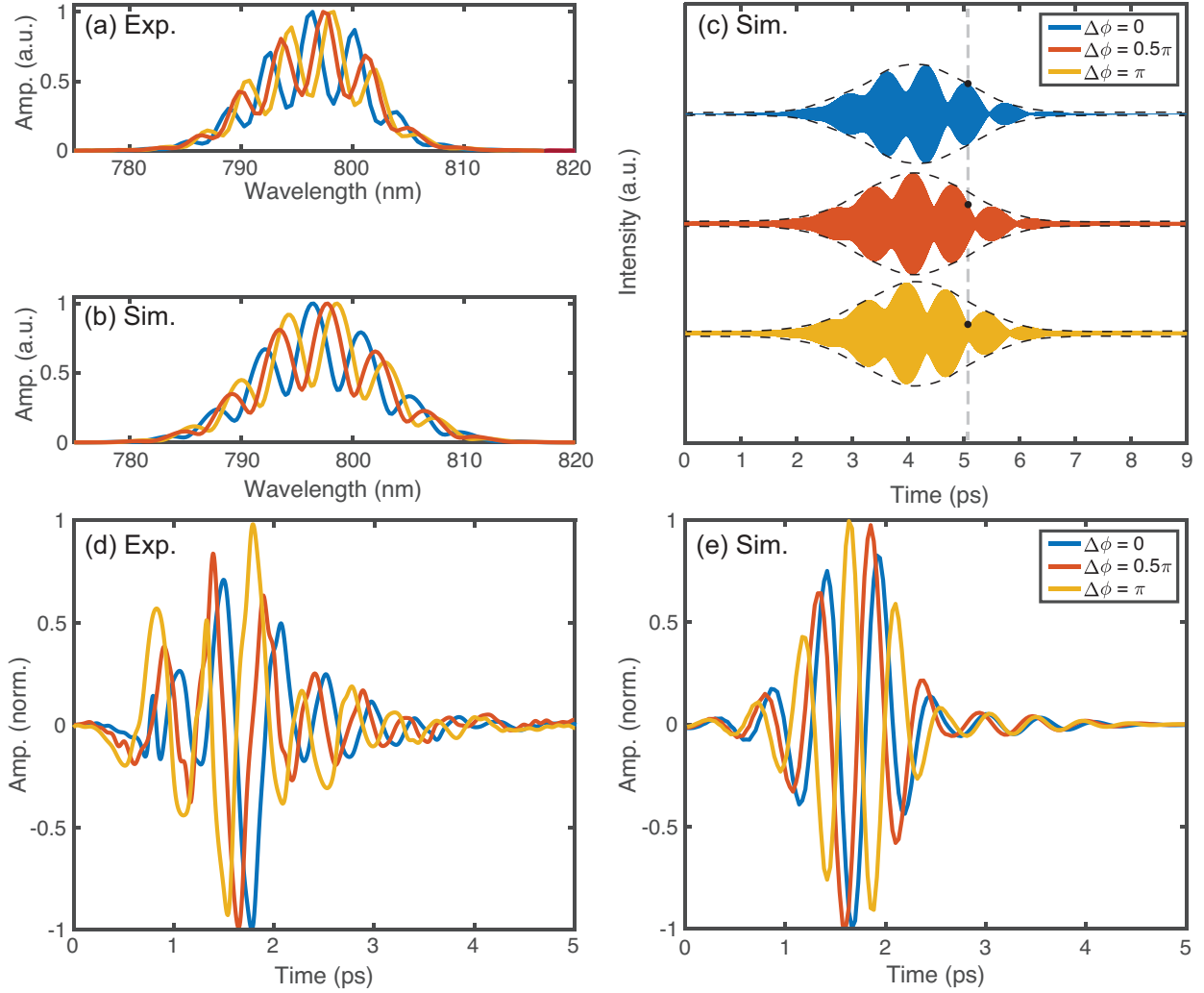}}
\caption{(a-b) Experimental and simulation results on spectra of Michelson interferometer output with $\Delta\phi$ set to 0,$\pi/2$ and $\pi$, respectively. (c) Reconstructed temporal profile of the pulse sequence obtained from the numerical calculation with the same set of $\Delta\phi$ values. (d) Representative experimental time-domain signals of the CEP-tunable THz radiation, frequency-down-converted from fundamental waves of various $\Delta\phi$. (e) Simulation results on THz waveforms for $\Delta\phi=0,\pi/2$ and $\pi$.}
\label{fig4}
\end{figure}

To demonstrate coherent control over CEP of narrowband THz pulse, a pair of BK7 wedges was inserted into the Michelson interferometer to introduce a relative difference in carrier-envelope phase ($\Delta\phi$) between the two stretched near-IR fundamental waves, which collectively displace the nodes of the beat-note-modulated global intensity profile of the output pulse sequence in the time domain. This inter-pulse CEP difference can be inferred from spectral interference in the frequency domain, Fig.\ \ref{fig4}(a), showing continuous shift in fringe pattern on the magnitude spectra upon varying the effective wedge-pair thickness. By the same token for demonstrating frequency tunability, we took the experimental spectrum of single pulse, computed the shifting interference spectrum on account of CEP difference $\Delta\phi$, shown in Fig.\ \ref{fig4}(b), and reconstructed the time-domain pulse sequence carrying a phase offset, covariant with $\Delta\phi$, between the beat note and intensity profile, shown in Fig.\ \ref{fig4}(c). All of the above attests the establishment of effective access to the beat-note phase of the fundamental wave by incorporation of the wedge pair, in a coherent manner. The coherence is utilized to periodically modulate ionization events and quiver motion of free electrons at the microscopic level so as to be pronounced in the THz radiation, at the final stage of freqeuncy down conversion. Indeed, we observed experimentally, Fig.\ \ref{fig4}(d), a one-to-one correspondence between CEP walk-off of the narrowband THz radiation and that carried by the fundamental wave. By displacing the wedge pair, we realized coherent control over CEP at three representative values, i.e., 0, $\pi$/2 and $\pi$, echoing the prediction of ADK-model-based simulation in Fig.\ \ref{fig4}(e).


Our results reported herein admittedly have discrepancies to a certain extent when comparing the theoretical prediction and experimental results, manifested by slight deformation in the waveform of THz emission and the limited duration. Ideally, a much wider tuning range and narrower bandwidth of the THz radiation should be feasible upon further improving the apparatus. Nevertheless, our experiment demonstrates a new method of generating CEP-controllable and center-frequency-tunable narrowband THz radiation by exploiting Michelson interferometer for coherent control of zero- and first-order inter-pulse phase difference. As a guide to future experiments, we expect increasing the power of the driving pulse sequence and applying DC bias for the facilitation of air-plasma filamentation can significantly improve the THz emission characteristics. We believe these improvements would produce intense narrowband THz radiation, which can serve as an ideal multi-cycle excitation beam for coherent manipulation of quantum materials at THz frequencies. The successful demonstration of the CEP control of the narrowband THz radiation establishes a direct mapping from the CEP of the input optical pulse to the output THz signal, implying a potential solution to CEP stabilization of the narrowband THz radiation. Conversely, the detection of the narrowband THz time-domain signal can also be used to probe the phase of the input optical pulse, furnishing a new CEP metrology.

In conclusion, by adopting the chirped-beating method, we have realized the generation of tunable narrowband THz radiation from air plasma induced by two-color laser pulse sequence. Most importantly, the coherent CEP control over the resultant narrowband THz radiation has been successfully demonstrated. It was achieved by tuning the zero-order phase difference of the twin pulses used for the synthesis of the periodically modulated pulse sequence, which in turn dictates the timing of modulation to the ionized electrons. Our work verified the feasibility of applying the chirped-beating method to two-color laser-induced air plasma scheme for the coherent manipulation of narrowband THz radiation characteristics.

\section*{Funding} Hong Kong Research Grants Council (Project No. ECS26302219, GRF16303721); National Natural Science Foundation of China Excellent Young Scientist Scheme (Grant No. 12122416); HKUST Research Equipment Competition (REC19SCR14), 2020-21 UROP Support Grant.

\bibliographystyle{ieeetr}
\bibliography{library}

\end{document}